\documentclass[conference]{IEEEtran}
\usepackage{cite}
\usepackage{amsmath}
\usepackage{enumitem}
\usepackage{amssymb}
\usepackage{amsthm}
\usepackage{graphicx,epstopdf}
\usepackage{multicol}
\usepackage{textcomp}
\usepackage{multirow}
\usepackage{flushend}
\usepackage{color}
\makeatletter
\newcommand*\bigcdot{\mathpalette\bigcdot@{0.85}}
\newcommand*\bigcdot@[2]{\mathbin{\vcenter{\hbox{\scalebox{#2}{$\m@th#1\bullet$}}}}}
\makeatother
\usepackage{lipsum}
\usepackage{cuted}

\usepackage{wrapfig, blindtext}

\begin{document}
\title{Optimal Sizing and Siting of Multi-purpose Utility-scale Shared Energy Storage Systems\vspace{-0.5ex}}

\author{Narayan Bhusal*, \emph{Student Member, IEEE}, Mukesh Gautam*, \emph{Student Member, IEEE}, \\Mohammed Benidris*, \emph{Member, IEEE}, and Sushil J. Louis**, \emph{Member, IEEE}, \\
*Department of Electrical and Biomedical Engineering, 
\\ **Department of Computer Science and Engineering,\\  
University of Nevada, Reno, NV 89557, USA\\
Emails: bhusalnarayan62@nevada.unr.edu, mukesh.gautam@nevada.unr.edu, \\mbenidris@unr.edu, and sushil@cse.unr.edu\vspace{-1ex}}

\maketitle
\thispagestyle{empty}
\pagestyle{empty}

\begin{abstract}
This paper proposes a nondominated sorting genetic algorithm II (NSGA-II) based approach to determine optimal or near-optimal sizing and siting of multi-purpose (e.g., voltage regulation and loss minimization), community-based, utility-scale shared energy storage in distribution systems with high penetration of solar photovoltaic energy systems. Small-scale behind-the-meter (BTM) batteries are expensive, not fully utilized, and their net value is difficult to generalize and to control for grid services. On the other hand, utility-scale shared energy storage (USSES) systems have the potential to provide primary (e.g., demand-side management, deferral of system upgrade, and demand charge reduction) as well as secondary (e.g., frequency regulation, resource adequacy, and energy arbitrage) grid services. Under the existing cost structure, storage deployed only for primary purpose cannot justify the economic benefit to owners. However, delivery of storage for primary service utilizes only 1-50\% of total battery lifetime capacity. In the proposed approach, for each candidate set of locations and sizes, the contribution of USSES systems to grid voltage deviation and power loss are evaluated and diverse Pareto-optimal front is created. USSES systems are dispersed through a new chromosome representation approach. From the list of Pareto-optimal front, distribution system planners will have the opportunity to select appropriate locations based on desired objectives. The proposed approach is demonstrated on the IEEE 123-node distribution test feeder with utility-scale PV and USSES systems. 
\end{abstract}
\IEEEpeerreviewmaketitle

\section{INTRODUCTION}
Energy storage systems have become essential components to modern distribution systems to overcome technical and operation challenges introduced by renewable energy sources (RESs). Also, falling prices of batteries and incentive programs have led to their wide adaption \cite{GREENBIZ}. California Public Utilities Commission has approved a new target---1.3 gigawatt (GW), out of which 200 megawatt (MW) on the customer-side needs to be installed in the state by 2020---for energy storage systems \cite{64987}. Large utilities, such as Southern California Edition (SCE) and Pacific Gas and Electric (PG \& E), have received approval from California public utility service to build storage facilities (195 MW for SCE and 567.5 MW for PG\&E) \cite{GREENBIZ}. Also, Arizona Public Service (APS) is planning to install 850 MW storage by 2025 \cite{GREENBIZ}. The economic justification is vital for these new installations of energy storages to be sustainable. Nevertheless, under the existing cost structure, economic benefits of storage systems that are deployed for primary purposes only (e.g., demand side management, deferral of system upgrade, and demand charge reduction) cannot be justified as these storages utilize only $1\%$--$50\%$ of total battery lifetime capacity \cite{RMI1541}.

Behind-the-meter (BTM) energy storage systems are expensive and not fully utilized due to the single-use and being dispersed on large geographical areas \cite{RMI1541}. Also, it is difficult to coordinate between a large number of BTM energy storage systems to provide grid services. On the other hand, utility-scale shared energy storage (USSES) systems have the potential to provide grid services and increase the utilization of photovoltaic (PV) systems and other RESs. They also provide opportunities for owners of distributed energy resources (DERs) to lease parts of the storage instead of buying individual energy storage systems.  

Development of multi-use strategies and models for community based USSES has become critical for both customers and utilities. Also, new multi-use business models which utilize battery storage for both primary and secondary (e.g., frequency regulation, resource adequacy, and energy arbitrage) services are needed to increase economic benefits for both USSES owners and investors \cite{RMI1541, ICPS2019NB}. Furthermore, determining optimal locations and sizes of USSES is an important factor to provide grid services. Optimal locations and sizes of energy storage system (ESS) can be used for several grid services such as improving power factor and serving demand for peak time period, improving voltage profiles and reducing the power loss, controlling high energy imbalance charges, and improving power quality and system reliability \cite{6397646}. 

Several studies have been conducted to assess the importance of storage systems in providing grid services. In \cite{6725376}, energy storage systems have been used for peak shaving. The authors of \cite{8027056} have utilized batteries for frequency regulation and peak shaving through a joint optimization framework. In \cite{6730958}, a battery storage system has been used to increase the utilization of RESs. A method to quantify the required energy storage to firm up wind generation has been proposed in \cite{7883840}. In \cite{7968249}, the authors have used battery storage to defer upgradation of distribution feeders. Optimal sizes and locations of battery storage units have been determined for voltage regulation in \cite{7286059}. Analysis of potential grid services that can be provided by USSES systems and potential technical and operation challenges have been discussed in \cite{RMI1541}. The economic viability of several battery storage technologies has been presented in \cite{DIVYA2009511, dunn2011electrical}. Benefits, applications, and technologies associated with utility-scale energy storage system have been provided in \cite{PURDUE1395}. Calculation of stacked revenue and technical benefits of a grid-connected energy storage systems has been presented in \cite{8334653}.

Numerous approaches have been proposed in the literature for optimal sizing and siting of distributed generation (DG) and energy storage systems (ESS). A review of models, methods, and future directions for distributed generation placement in distribution systems have been provided in \cite{6418071}. A two-stage sequential Monte Carlo simulation (MCS)-based stochastic strategy has been proposed in \cite{PES2020NB} to determine the minimum size of movable energy resources (MERs) for service restoration and reliability enhancement. Genetic algorithm (GA) and particle swarm optimization (PSO) have been combined together in \cite{MORADI201266} for the optimal sizing and siting of DGs to minimize power losses, improve voltage regulation, and enhance voltage stability. A GA based framework has been presented in \cite{6400274} for the optimal placement of ESS in a high wind integrated systems to minimize the operational cost. In \cite{6868308}, multi-objective combined PSO and non-dominated sorting genetic algorithm (NSGA-II) based approach has been proposed for optimal siting of ESS to minimize the operational cost and improve the voltage profile. Authors of \cite{6476054} have proposed a GA based approach for optimal planning of ESS in smart grids to consider the cost sustained by asset owners through total planning period.

Although several approaches have been proposed for sizing and siting of ESSs, they focus on very large scale ESS for transmission level system and for DG placement, which has different operational characteristics than USSES systems. Also, most of the literature have combined all the objective functions to develop a single objective with various arbitrary chosen weights. Therefore, the proposed work is novel in the sense that it determines the optimal sizes and sites of multi-purpose USSES with leasing opportunities at the distribution level. Also, this paper emphasizes on the concept of multi-purpose community based USSES for the economic justification. In terms of the GA representation, the approach presented in this paper disperses the resources rather than placing them in a single location. Dispersing resources have several benefits specially during the time of extreme disasters (natural disasters as well as man-made attacks) \cite{8966351}.

This paper proposes an NSGA-II based approach to determine the optimal or near optimal sizes and locations of USSES for multiple purposes (voltage deviation and loss minimization). For each candidate set of locations, the contribution of USSES systems to grid voltage deviations and power losses are evaluated and diverse Pareto-optimal front is created. From the list of Pareto-optimal front, distribution system planners will have opportunity to select the appropriate sizes and sites based on the desired objectives. The proposed approach is demonstrated on the IEEE 123-node distribution test feeder with utility-scale PV and USSES systems. 

The rest of the paper is organized as follows. Section \ref{problem_formulation} discusses the problem of optimal sizes and sites of energy storage systems and their importance to grid services. Section \ref{solution_approach} presents the development of the proposed approach. Section \ref{cases} examines the proposed approach through case studies on IEEE-123 node test system. Finally section \ref{conclusion} provides concluding remarks.

\section{Problem Formulation} \label{problem_formulation}
Determining optimal sizes and locations of USSES systems can be determined based on several factors such as minimization of operating costs, power losses and voltage deviations, improvement of power quality and reliability, and frequency regulation. Power loss and voltage deviation minimization are two important operational measures which have significant impact on both technical and economic aspects of distribution system operation. Therefore, proper consideration should be given for minimizing power losses and voltage deviations while determining optimal sizes and sites of community based USSES. 

Power loss between bus $j$ and $k$ with photovoltaic (PV) at node $k$ can be computed as follows \cite{6205640}.
\begin{gather*}
           P_{Loss}=R_{jk}\frac{(P_{j}^2+Q_{j}^2)}{V_{j}^2}+ \\  \frac{R_{jk}}{V_{j}^2}(P_{PV}^2+Q_{PV}^2-2P_{j}P_{PV}-3Q_{j}Q_{PV})\left( \frac{G}{L}\right)\mbox{,} \tag{1}
            \label{equ:powerloss}
\end{gather*} 
where $P_j$ and $Q_j$ are the injected real and reactive power at bus $j$; $R_{ij}$ is resistance of line segment $i$ to $j$; $V_j$ is the voltage at node $j$; $P_{PV}$ and $Q_{PV}$ are, respectively, the real and  reactive power produced by a PV system at bus $k$; $G$ is  the  distance  from  a  bus with a PV system ($k$) to the source; and $L$ is the total feeder length from the source to bus $j$. 

The total power loss of the feeder at the $i$th hour can be calculated as follows. 
\begin{gather*}
    P_{T, Loss}^i=\sum_{j=1}^{N} P_{Loss}^i(j, j+1)\mbox{,} \tag{2}
    \label{equ:power loss total}
\end{gather*}
where $N$ is the total number of line segments.

Since the work presented here is for a planning purpose, the average of total power losses for a year is considered, which can be expressed as follows.
\begin{gather*}
    P_{avg, Loss}=\frac{\sum_{i=1}^{8760} P_{T, Loss}^i}{8760}\mbox{,} \tag{3}
    \label{equ:total average power loss}
\end{gather*}

Similar to average power losses, voltage deviations can also be averaged over a year as follows.
\begin{gather*}
    \Delta V_{avg}=\frac{ \sum_{i=1}^{8760}\Big(V_{max}^i-V_{min}^i\Big)}{8760}\mbox{,} \tag{4} \label{equ:voltage_dev}
\end{gather*}
where $V_{max}^i$ and $V_{min}^i$ are, respectively, the maximum and minimum system voltage at hour $i$.

Therefore, the objective function considered in this paper can be expressed as follows.
\begin{gather*}
                F1:\mbox{Minimize }P_{T, Loss}\mbox{,} \tag{5} \label{equ:obj_power_loss}\\
               F2:\mbox{Minimize } \Delta V_{avg}\mbox{,} \tag{6} \label{equ:voltage_deviation}
\end{gather*}
Subject to: 
\begin{gather*}
\sum S_{G,i}-\sum S_{L, i}-TL_{s}=0\mbox{,} \tag{7} \label{equ:gen_load}  \\ S_G^{min} \leq S_G \leq S_G^{max}\mbox{,} \tag{8} \label{equ:gen_limit} \\ S_{ij}\leq S_{ij}^{max}\mbox{,} \tag{9} \label{equ:loading_limit} \\ V_k^{min}\leq V_{k}\leq V_k^{max}\mbox{,} \tag{10} \label{equ:voltage_limit} \\ B_{SOC}^{min}\leq  B_{SOC}\leq B_{SOC}^{max}\mbox{,} \tag{11}\label{equ:BSOC_limit}
\end{gather*} 
where \eqref{equ:gen_load} denotes the power balance equation ($S_{G, i}$, $S_{L, i}$, and $TL_s$ represents the generation, load, and transmission loss, respectively); \eqref{equ:gen_limit} refers to the generation limits constraint; \eqref{equ:voltage_limit} represents voltage limits constraint; and \eqref{equ:BSOC_limit} denotes battery state-of-charge limits constraint.   

Since community based USSES systems are installed at the distribution level, which have several candidate locations, the problem becomes very challenging. For larger systems, the number of possible scenarios becomes dramatically very large and more specifically when multiple community based USSES units are needed. The complexity of the problem with fixed sizes and variable locations can be demonstrated using \eqref{equ:combination}.
\begin{gather*}
 S=\frac{L_{c}!}{n_{B}!(L_{c}-n_{B})!}\mbox{,}\tag{12}
\label{equ:combination}
\end{gather*}
where $L_c$ is total number of candidate nodes in a feeder and $n_B$ is the number of USSES systems to be installed at the feeder. 
  
In this work, we have considered both optimal sizing and siting of USSES systems, which makes the problem more complicated. For example, to place $10$, $15$, or $20$ USSES systems in $123$ possible locations, the total number of combinations are, respectively, more than $1.5\times10^{14}$, $7.01\times 10^{18}$, and $5.04\times 10^{22}$. As this is a planning problem, to properly incorporate the load changing scenario, all simulation are run for a year with one hour time steps. Therefore, exhaustive search could take months find the optimal and complete solution. Therefore, in this work, GA has been adopted to find the optimal or near optimal locations and sizes for community-based USSES systems to minimize the power losses and voltage deviations.
  
Several methods in the literature have used arbitrary chosen weighting factors to convert multi-objective problems into a single objective function. Also, the problem must be solved for every change in the desired priority in weight related considerations. To deal with these problems, non-dominated sorting genetic algorithm has been proposed in \cite{996017} which provides pareto-optimal fronts rather than one solution. The solution can be chosen from the pareto-optimal solutions based on desired objectives. NSGA-II has been proven to be most effective for multi-objective optimization on a number of power system benchmark problems \cite{5262957, 4359250, 7422916}. In \cite{5262957}, NSGA-II has been adopted to solve generation expansion planning problems. Electric distribution service restoration (minimize out-of service area, minimize switching operation, minimize power loss) has been performed using NSGA-II in \cite{4359250}. Authors of \cite{7422916} have adopted NSGA-II for optimal DG siting and sizing with storage systems and feeder reconfiguration effects.  

\section{Solution Approach} \label{solution_approach}
\subsection{Unbalanced Power Flow and Simulation Environment}
In this work, MATLAB and OpenDSS are integrated to perform the proposed work. In MATLAB, all the control commands and GA functions are performed which calls OpenDSS engine to perform unbalanced three-phase power flow. OpenDSS provides all the monitored information back to the MATLAB to perform the remaining tasks. The OpenDSS is an open source power system simulation environment for distribution system simulation, which is developed by Electric Power Research Institute (EPRI) \cite{EPRIOpenDSS}. The OpenDSS calculates unbalanced power flow using Newton's Method (note that Newton's Method implemented in OpenDSS is different from the Newton-Raphson method). 

\subsection{Representation of the Proposed Approach}
\begin{figure}
\vspace{-3ex}
    \includegraphics {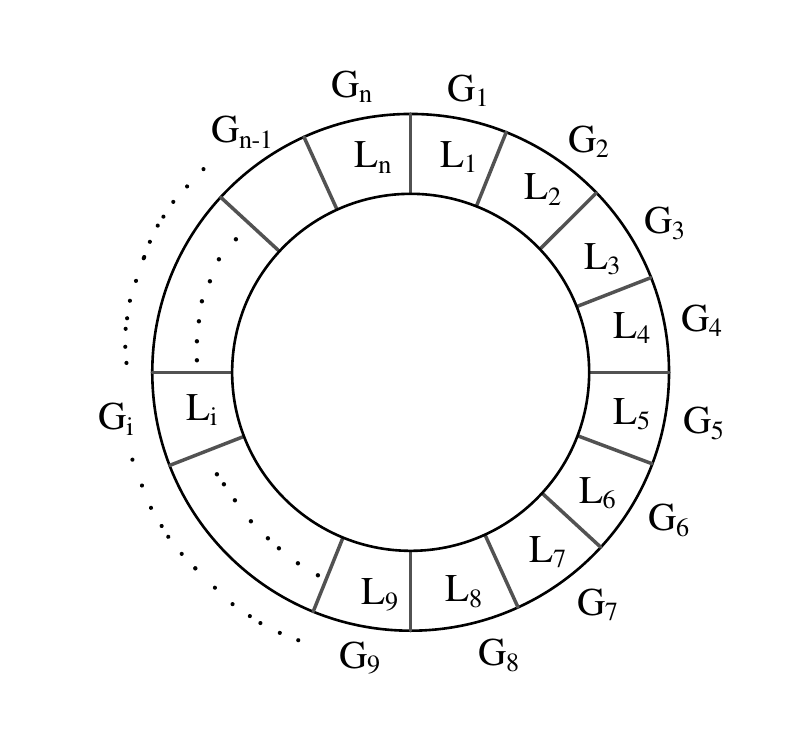}
    \vspace{-3ex}
    \caption{Representation of chromosome to avoid multiple USSES systems in a single location}
    \label{fig:representation}
\end{figure}

In the proposed problem, the USSES need to be dispersed in different locations rather than placing them in one single location; thus, conventional representation techniques can not perform this task. Therefore, we propose a novel representation technique as shown in Fig. \ref{fig:representation},  where for $N$ possible nodes, $G_1$, $G_2$, $G_3$ .... $G_i$...$G_{n-1}$ and $G_n$ are node numbers from $1$ to $N$. The locations obtained from these selections may not be unique; therefore, unique locations are obtained using GA numbers ($G_1$, $G_2$, $G_3$ .... $G_i$...$G_{n-1}$, $G_n$) in terms of $L_1$, $L_2$, $L_3$, $L_4$, ... $L_i$, .., $L_{n-1}$, and $L_n$. The unique locations are created based on the distance between two GA numbers. In this process, all newly created unique numbers are checked for similarity and if they are similar, the distance between a particular number and the next number in clockwise direction is determined. This process is repeated until the chromosome becomes completely unique. For example, if $[x_1, x_2, x_3, x_4, x_5]=[G_1, G_3, G_12, G_3, G_{43}]$ is a chromosome obtained after performing selection. The unique location can be obtained as follows. The first term is obtained as $x_1^{'} = G_1$, and rest of the terms are obtained from the distance between them using the technique shown in Fig. \ref{fig:representation}. For example, second term $x_2^{'}$ is distance between the number $x_2$ and $x_1$. $x_2^{'}$ is compared with all the elements of the newly created chromosome (for $x_2^{'}$, here it has to compare with $x_1^{'}$) and if it is similar the distance between $x_1$ and $x_2$, the distance between $x_2$ and $x_3$ are added together to create the unique the number. This process is repeated until all the numbers in the newly created chromosome are unique. The limitation of this representation is when chromosome is $[G_n, G_n, G_n, G_n, G_n]$, it becomes never ending loop, therefore, for this case a random number is generator (less than $G_n$) and one of the number of the chromosome is replaced with the random number.

\section{Case Studies and Results}\label{cases}
\begin{figure}
    \hspace{-4ex}
    \includegraphics[scale=0.8]{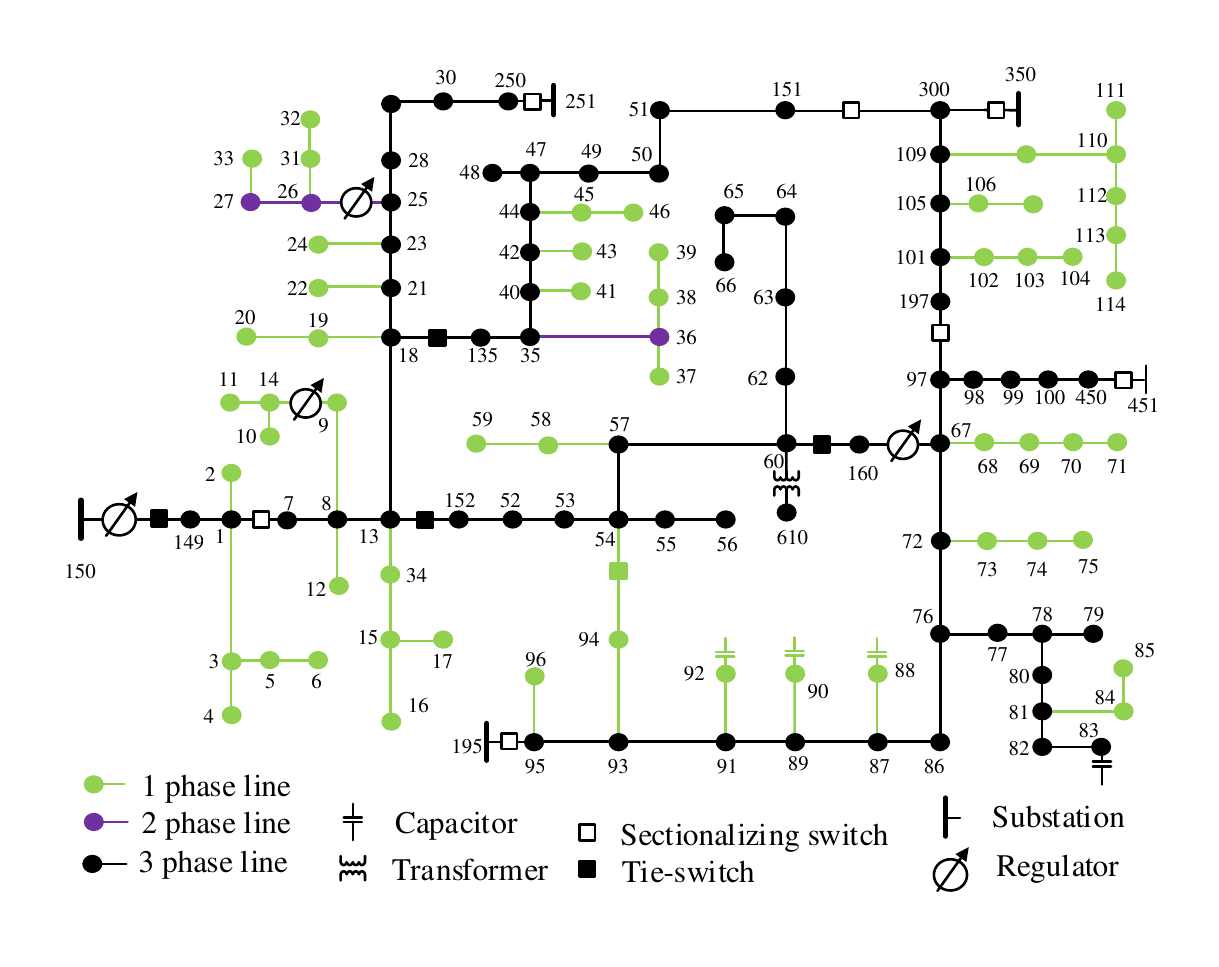}
    \caption{IEEE 123 node distribution test system}
    \vspace{-2ex}
    \label{fig:IEEE123}
\end{figure}

In order to validate the proposed approach, all simulations have been performed on the IEEE-123 bus radial distribution test feeder\cite{6960676} using MATLAB and OpenDSS integrated environment.  The IEEE-123 node test feeder, as shown in Fig. \ref{fig:IEEE123}, is characterized by having overhead and underground lines, four voltage regulators, four shunt capacitor banks, multiple sectionalizing and tie-switches, and unbalanced loading with constant current, power, and impedance models. The total real and reactive loads of this system are, respectively, $3490$ kW and $1925$ kVar. Network data of the IEEE 123-node test feeder are given in \cite{IEEEFEEDERS}. Ten PVs each of sized $500$ kVA ($450$ kW at maximum power point tracking) are placed at nodes $7$, $13$, $25$, $3$, $47$, $56$, $62$, $72$, $82$, and $105$.

\begin{figure}
    \centering
    \includegraphics[scale=0.44]{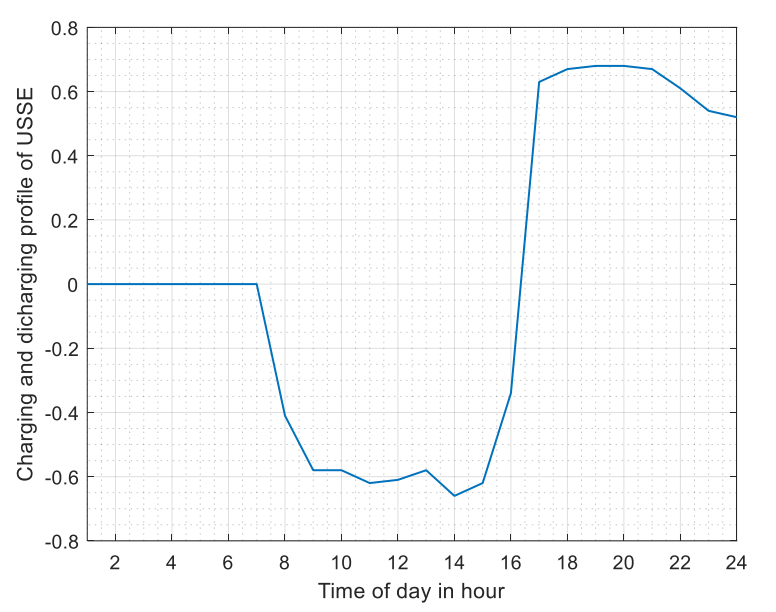}
    \caption{USSES profile developed using the combine profile of PV output and load demand.}
    \vspace{-2ex}
    \label{fig:USSES_profile}
\end{figure}

In this paper, it is assumed that community based USSES units are fully charged and discharged in every $24$ hours (as in \eqref{equ:battery}) to ensure maximum utilization of batteries.
 \begin{gather*}
     \sum_{t=1}^{24} P_{Batt}^{t}=0, \tag{13} \label{equ:battery}
 \end{gather*}
 where $P_{Batt}$ is charging (negative) or discharging (positive) power of a battery at any hour $t$ of a day. Charging and discharging profiles of community based USSES is drived based on expected energy produced from solar PV and load demand as shown in Fig. \ref{fig:USSES_profile}.

\begin{figure}
    \includegraphics[scale=0.45]{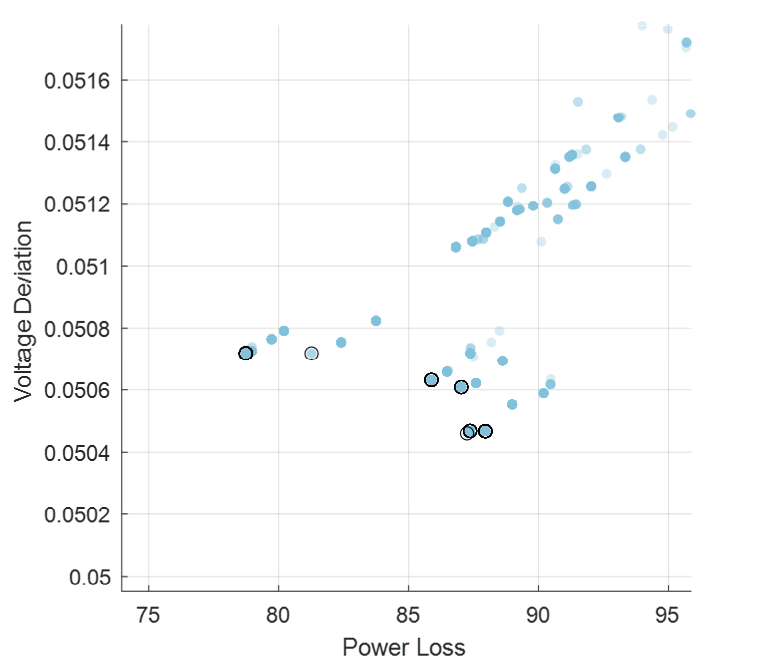}
    \caption{Pareto optimal solution, respective sizes and cites are as shown in Table \ref{tab:pareto_solution}}
    \label{fig:pareto-front}
\end{figure}

The optimal sizes and sites of $5$ community based USSES are searched on the IEEE-123 feeder. For siting of USSES, out of $123$ nodes, only $56$ three phase nodes are selected (all three phase nodes are represented by black dot in Fig. \ref{fig:IEEE123}). For sizing of USSES,  sizes ranging from $100$ to $1000$ kWh are considered. The parameters used for performing the NSGA-II are as follows: two point mutation (one for size and one for site), two point crossover, binary representation with total string length of $80$ ($30$ for site and $50$ for sizing), $50$ population size, and $80$ generations (population size and number of generations are determined after some trial).  The obtained Pareto-optimal solutions are as shown in Table \ref{tab:pareto_solution} and the plot between two objectives (Voltage deviation and Power loss) is as shown in Fig.\ref{fig:pareto-front}. The simulations are run for $10$ times (this number can be any number, we just checked for $10$ times, every time result were similar, therefore, we didn't check further), in every simulation runs the sizes and sites obtained for each cases are similar. From this we can draw the conclusion that the obtained solution is good set of Pareto-optimal solution. 
\begin{table*}
    \caption{Size and sizing of USSES and respective power loss and voltage deviation.}
    \vspace{-15ex}
        \hspace{-14ex}\includegraphics[scale=1]{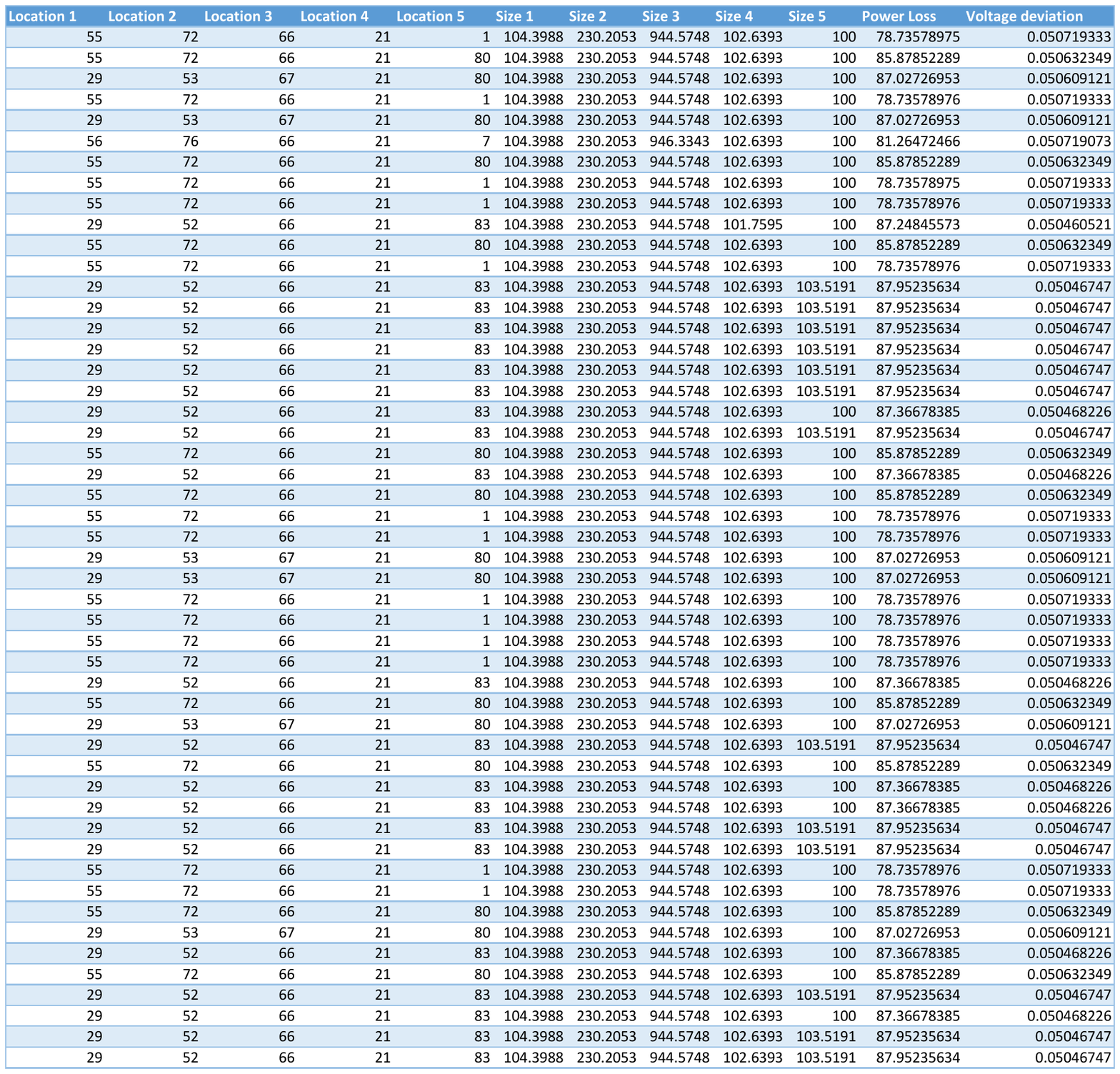}
        \vspace{-75ex}
        \label{tab:pareto_solution}
\end{table*}

\section{Conclusion}\label{conclusion}
This paper has proposed an NSGA-II based approach to determine optimal or near-optimal sizing and siting of community based USSES system in solar photovoltaic integrated distribution system for multiple-purpose (voltage deviation minimization, and loss minimization). For each candidate set of locations and sizes, the contribution of USSES systems to grid voltage deviation and power loss were evaluated and diverse Pareto-optimal front were created. From the list of Pareto-optimal front, distribution system planners will have opportunity to select the appropriate location based on desired objective. The proposed approach was demonstrated on the IEEE-123 node distribution test feeder with utility-scale PV systems. Similar results in multiple runs validated the optimality of the obtained Pareto-optimal solution.

\bibliographystyle{IEEEtran}
\bibliography{References.bib}
\end{document}